\documentclass[epj,referee]{svjour}
\usepackage{graphics}
\begin{document}
\title{Lifetime of heavy hypernuclei and its
implications on the weak $\Lambda N$ interaction
}
\author{W. Cassing\inst{1}
\and L. Jarczyk \inst{2}
\and B. Kamys \inst{2}
\and P.Kulessa \inst{3,4}
\and H. Ohm \inst{3}
\and K. Pysz \inst{3,4}
\and Z. Rudy \inst{2,3}
\and O.W.B. Schult \inst{3}
\and H. Str\"oher \inst{3}
}                     
%
%
\institute{Institut f\"ur Theoretische Physik, Justus Liebig Universit\"at Giessen,
D-35392 Giessen, Germany
\and M. Smoluchowski Institute of Physics, Jagellonian University,
PL-30059 Cracow, Poland
\and Institut f\"ur Kernphysik, Forschungszentrum J\"ulich,
D-52425 J\"ulich, Germany
\and H. Niewodnicza\'nski Institute of Nuclear Physics, PL-31342
Cracow, Poland }
\date{Received: date / Revised version: date}
%
\abstract{
 The lifetime of the $\Lambda$--hyperon in heavy
hypernuclei  -- as measured by the COSY--13 Collaboration  in
proton -- Au, Bi and U collisions at COSY--J\"ulich -- has been
analyzed to yield $\tau_{\Lambda}= (145 \pm 11$) ps. This value
for $\tau_{\Lambda}$ is compatible with the lifetime extracted
from antiproton annihilation on Bi and U targets, however, much
more accurate. Theoretical models based on the meson exchange
picture and assuming the validity of the phenomenological $\Delta
I$=1/2 rule predict the lifetime of heavy hypernuclei to be
significantly larger (2 -- 3 standard deviations). Such large
differences may indicate that the assumptions of these models are
not fulfilled. A much better reproduction of the lifetimes of
heavy hypernuclei is achieved in the phase space model, if the
$\Delta I$=1/2 rule is discarded in the nonmesonic $\Lambda$
decay.
} 
\maketitle
\section{Introduction}
\label{sec:intro}

The $\Lambda$ hyperon decay can be studied for free hyperons as
well as for hyperons colliding with nucleons inside the nuclear
medium. In the first case it proceeds via the mesonic process,
$\Lambda \rightarrow \pi$ +N, with an energy release of about 38
MeV, whereas collisions with nucleons lead to the nonmesonic
decay, {\it e.g.} N + $\Lambda \rightarrow$ N + N, with an energy
release of ($\sim$ 180 MeV).

The mesonic decay also occurs for hyperons bound in hypernuclei,
but it is  strongly inhibited for all but the lightest hypernuclei
due to Pauli blocking of the nucleon final states. The nonmesonic
decay, on the other hand, can be  studied only in hypernuclei
because neither $\Lambda$ hyperon beams nor targets are available.
Due to the immense difficulty in producing $\Lambda$ hypernuclei
and of subsequently detecting their decay the available
experimental  data on the nonmesonic process are scarce and have
large uncertainties.

Most of the measurements for the decay of hypernuclei have been
based on limited statistics and been predominantly performed for
light hypernuclei (see {\it e.g.} the reviews
\cite{COH90A,OSE98,ALB01} or refs.
\cite{MON74,GRA85,SZY91,NOU95,BHA98,PAR00,HAS02}). Even the total
decay rate (or inverse lifetime) of heavy hypernuclei was up to
very recently known only with a large error \cite{ARM93}. The
experimental knowledge of the partial decay rates is also not
satisfactory, {\it e.g.} the experimental studies devoted to light
(A $\le$ 28)\cite{MON74,GRA85,SZY91,NOU95,HAS02} and medium heavy (40 $<$
A $<$ 100) hypernuclei \cite{LAG64,CUE67,GAN67} report different
values for the neutron and proton induced $\Gamma_n/\Gamma_p$
decay rates. The results for light hypernuclei are close to unity
whereas those for heavy hypernuclei vary between 1.5 and 9.0. The
experimental situation -- together with uncertainties in the
theoretical description -- show that the nonmesonic process is
barely understood so far.

We recall that in the Standard Model the weak $\mid \Delta S \mid
= 1$ transitions can proceed with both $\Delta I =1/2$ and $\Delta
I = 3/2$ amplitudes.  However, it was found experimentally (in the
decays of free kaons and hyperons) that the $\Delta I = 1/2$
amplitudes dominate by far the $\mid \Delta S \mid = 1$
non-leptonic weak interactions \cite{DON86}. This suppression of
the $\Delta I = 3/2$ amplitude was explained by Miura --
Minanikawa  \cite{MIU67} and Pati -- Woo \cite{PAT71} in terms of
the colour symmetry of the valence quarks in the baryon. Thus, one
is tempted to assume a dominance of $\Delta I = 1/2$ transitions
also in the nonmesonic decay of the $\Lambda$ - hyperon. It was,
however,  observed that  theoretical calculations involving this
assumption -- i.e. only $\Delta I = 1/2$ transitions --
systematically underpredict  the ratio $\Gamma_{n}$/$\Gamma_{p}$
of nonmesonic decay rates induced by neutrons ($n+\Lambda
\rightarrow n + n$) to the decay rates induced by protons
($p+\Lambda \rightarrow p + n$) \cite{HAG00}. Several attempts
have been made to reconcile this discrepancy {\it e.g.} in Refs.
\cite{PAR98,ALB91,RAM97,DUB96,SAV96,SAS00}, but none of them has
solved this problem in a convincing way.

This leads  to the conclusion that the contribution of the $\Delta
I = 3/2$ transition to the nonmesonic decay of the $\Lambda$
hyperon might  not be negligible, {\it i.e.} the $\Delta I = 1/2$
rule should be violated \cite{DOV87,COH90,SCH92,RUD99,ALB00}. The
arguments presented in favor of this hypothesis in  refs.
\cite{DOV87,COH90,SCH92,ALB00} have been based essentially on the
observed nonmesonic decay widths of the lightest hypernuclei.
However, the experimental uncertainties are too large to allow for
any definite conclusion. It is thus necessary to get information
on the (possible) violation of the $\Delta I = 1/2$ rule from
other properties of hypernuclei, {\it e.g.} from the mass
dependence of the lifetime of hypernuclei  as addressed in ref.
\cite{RUD99}.

As far as experiments are concerned it can be stated from the
inspection of Table \ref{tab:1}, that the data - with exception of
the experiment performed with an $e^-$ beam in Kharkov
\cite{NOG86} - agree within the limits of errors. In ref.
\cite{KUL98} it has been shown, that a hypernucleus fraction
decaying on a timescale of 2700 ps (as quoted in \cite{NOG86})
must be smaller by orders of magnitude compared to the fraction of
hypernuclei decaying on timescales of 200 ps. However, the errors
for $\tau_\Lambda$ in the measurements from \cite{BOC86,ARM93} are
so large, that no severe constraints could be imposed on the
various theoretical models for the nonmesonic decay.


\begin{table}[h]
\caption{The lifetimes of heavy hypernuclei from $e^-$ and
$\bar{p}$ induced reactions from refs.
\protect\cite{NOG86,BOC86,ARM93}. The numbers given in parenthesis
represent the systematic errors. }
\label{tab:1}       
\begin{tabular}{lcll}
\hline\noalign{\smallskip} Target         & $\tau_{\Lambda}$ / ps
& Ref. & Comment \\
\& projectile  &                       & &
\\ \noalign{\smallskip}\hline\noalign{\smallskip} Bi + e & 2700
$\pm$ 500         & \cite{NOG86} &  \\
Bi + $\overline{p}$ &
250$^{+250}_{-100}$    & \cite{BOC86} &  \\
Bi + $\overline{p}$ &
180$\pm$ 40 ($\pm$ 60) & \cite{ARM93} & Reanalysis\\
                    &                        &              &
of data from \cite{BOC86}\\ U + $\overline{p}$ & 130 $\pm$ 30
($\pm$ 30)& \cite{ARM93} &  \\
\noalign{\smallskip}\hline
\end{tabular}
\end{table}


In order to improve the situation,   experiments with $Au, Bi$ and
$U$ targets have been performed during the last years at the
Forschungszentrum J\"ulich using the internal proton beam of the
COSY accelerator by the COSY--13 Collaboration. We briefly
describe the different stages of the proton-nucleus reactions --
leading to hypernucleus formation and their delayed fission due to
the $\Lambda N \rightarrow NN$ decay -- in section 2. The
experimental setup used to distinguish prompt and delayed fission
events is sketched in section 3 and an overview of the
experimental results for the $\Lambda$ lifetime is given in
section 4. In section 5 we discuss the implication of the latter
results for the selection rule $\Delta I = 1/2$ in the $\Lambda N
\rightarrow NN$ transition. A discussion of open problems and a
summary are presented in sections 6 and 7, respectively.

\section{Heavy hypernuclei formation in $p+A$ reactions and their decays}
\label{sec:exper}
In case of heavy hypernuclei the application of direct timing
methods - as adopted for light hypernuclei  - is not feasible due
to the large background of light particles produced. This problem
is circumvented by detecting heavy fragments from the fission
processes, which are induced by the $\Lambda$--hyperon decay in
heavy hypernuclei. The technique used is the recoil shadow method
originally suggested by Metag { et al.} for the measurement of
fission isomers \cite{MET74}. It has also been employed by Armstrong et
al. \cite{ARM93} in the lifetime measurements with antiprotons.

A novel approach to produce heavy hypernuclei for lifetime
measurements -- as performed by the COSY--13 Collaboration -- is
to use proton collisions on heavy targets like $U, Bi$ or $Au$.
The possibility to vary the beam energy  allows to measure the
background (at a low beam energy, e.g. of 1 GeV) concurrently with
the effect (e.g. at 1.9 GeV) by operating COSY in a supercycle,
which has not been possible in the $\bar{p}$ induced reactions in
\cite{ARM93}. Furthermore, a variation of the projectile energy in
proton induced reactions permits to find out whether an ordinary
fission isomer might fake the decay of a hypernucleus. Such a test
is also not possible in antiproton--nucleus interactions since the
center-of-mass energy is fixed for  stopped antiprotons and always
above threshold for $\Lambda$ production. Furthermore, in $p + A$
reactions a large part of the proton momentum is transferred to
the hypernucleus  such that the surviving hypernuclei move faster
than in $\bar{p}$ induced reactions; this increases   the
sensitivity of the recoil shadow method for lifetime measurements
accordingly.

\begin{figure}
\resizebox{0.5\textwidth}{!}{%
  \includegraphics{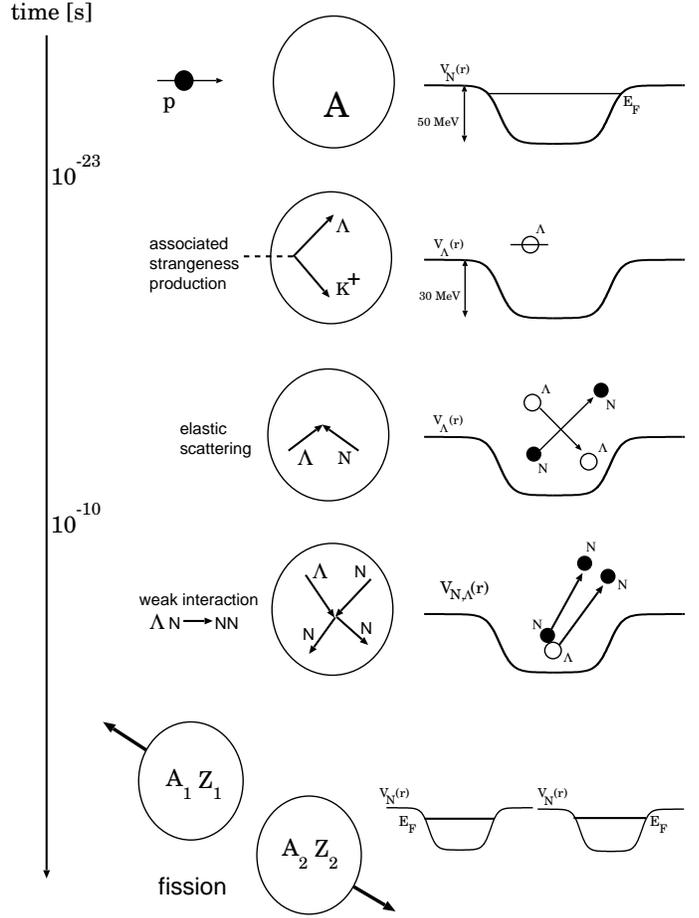}}
 \caption{Time evolution of a proton-nucleus collision from i) the
 initial configuration up to v) the delayed
fission of a hypernucleus: ii) associated strangeness production;
iii) elastic $\Lambda$N scattering; iv) decay of a $\Lambda$
hyperon via the $\Lambda$N $\rightarrow$ NN process leading v) to
fission of the excited nucleus. The right part shows the nucleon
potentials during the various phases.} \label{fig1}
\end{figure}

For illustration we show in Fig. \ref{fig1} the various stages and
time scales involved in the $p+A$ reaction from i) the initial
configuration to ii) the associated hyperon production in the
target nucleus by $pN$ inelastic scattering ($\sim 10^{-23}s$),
iii) $\Lambda$ hyperon capture in the residual nucleus via elastic
$\Lambda N$ scattering ($\sim 10^{-22}s$), iv) the $\Lambda N
\rightarrow NN $ reaction on the time scale of 200 $ps$ leading to
v) delayed fission of the hypernucleus. The right part shows the
nucleon potentials during the various phases.


Due to the complexity of these reactions the various stages
illustrated in Fig. \ref{fig1} have been simulated by
coupled-channel Boltzmann-Uehling-Uhlenbeck (CBUU) transport
calculations for the fast nonequilibrium phase
\cite{WOL90,WOL93,RUD95,RUD96} followed by Hauser-Feshbach
calculations for the statistical evaporation phase \cite{GAV93}.
The transport model employed has been used for a variety of
hadron-nucleus and nucleus-nucleus reactions from low to
relativistic bombarding energies and been tested with respect to
the overall reaction dynamics as well as the production of strange
and nonstrange hadrons (for reviews see refs. \cite{CAS90,CAS99}).
The CBUU calculations provide information on i) the formation
cross section of 'hot' hypernuclei as well as on ii) the
properties of the hypernuclei produced -- i.e. primary mass,
charge, excitation energy, linear momentum, angular momentum etc.
-- in a given reaction. The latter information from the CBUU
calculation then is used to evaluate (within Hauser-Feshbach
calculations)  for each event the subsequent statistical decay as
well as the probability $P_s$ of a heavy hypernucleus to survive
in competition with prompt fission \cite{RUD96}. Thus, the final
distribution in mass and charge of the 'cold' hypernuclei --
reached after $\sim 10^{-18}s$ (see below) -- is evaluated
together with their individual ($A,Z$ dependent) velocity
distribution in the laboratory frame. The probability for delayed
fission $P_{f_\Lambda}$ -- as induced by the $\Lambda N
\rightarrow NN$ reaction for a hyperon from the $S$-state in the
individual hypernuclei on a timescale of 200 $ps$ -- is calculated
again within the Hauser-Feshbach formalism \cite{RUD96}. The
kinematics of the fission fragments, furthermore, is simulated
according to the Viola systematics \cite{VIO85} assuming isotropic
angular distributions for the fission fragments in the rest frame
of the decaying hypernucleus. For details we refer the reader to
refs. \cite{RUD95,RUD96,RUD98}.

\begin{figure}
\resizebox{0.5\textwidth}{!}{%
  \includegraphics{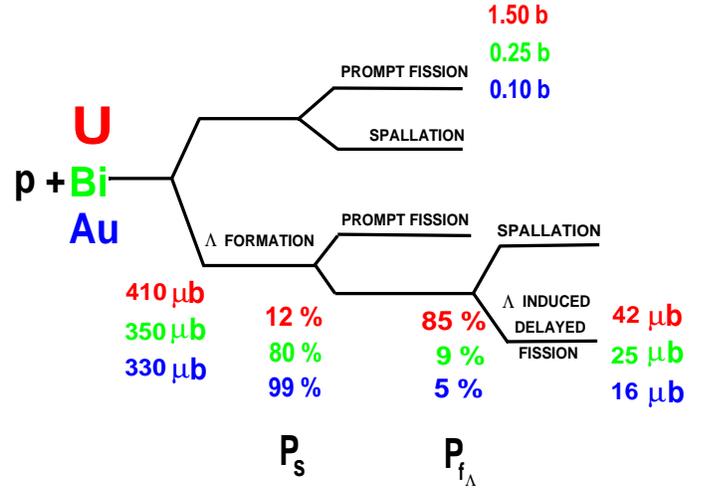}
} 
\caption{ Schematic representation of contributions from different
competing processes in p+Au, p+Bi, p+U reactions at T$_p$=1.9 GeV
according to the CBUU + Hauser-Feshbach calculations (see text).
The experimental cross sections for prompt fission have been taken
from Refs. \protect\cite{HUD76,VAI81}.} \label{fig2}
\end{figure}

The cross sections for $Au, Bi$, and $U$ targets at $T_{lab}$ =
1.9 GeV -- as calculated from the CBUU + evaporation calculations
-- are displayed in Fig. \ref{fig2}, where we show the predicted
cross sections and branching ratios for all targets. The experimental
cross sections quoted in Fig. \ref{fig2} for prompt fission
have been taken from refs. \cite{HUD76,VAI81}. In contrast
to the large differences in the  prompt fission cross
sections, which amount to a factor of $\sim$ 15 for $U$  and $Au$
targets, the cross sections for delayed fission ($\sim$ 42, 25 and
16 $\mu b$ for $U, Bi$ and $Au$, respectively) are rather similar.
This is due to the fact that the probability to observe the
delayed fission of hypernuclei is determined by a product of two
probabilities: the survival probability P$_s$ of ('hot')
hypernuclei against  prompt fission and the probability
P$_{f_{\Lambda}}$ for fission of ('cold') hypernuclei  induced by
a $\Lambda$ - hyperon decay. These two probabilities correspond to
opposite processes; their sum is approximately equal to unity. We
find, furthermore, that also their product remains constant within
a factor of 2$-$3.


The comparison of the cross sections for delayed fission of
hypernuclei and prompt fission of target nuclei in Fig. \ref{fig2}
shows that in experiments with  $Bi$ and $Au$ targets  the
same statistics for delayed fission fragments can be obtained using 2 -- 3 times
the beam time for a corresponding uranium experiment.  On the other
hand, the background from prompt fission in the $Bi$ or $Au$ experiments
is much smaller because the ratio of the
prompt to the delayed fission cross sections is small compared to
a $U$ target. This reduces the load on the detectors in the prompt
fission region for $Au$ and $Bi$ targets by about an order of
magnitude relative to $U$. These expectations (calculations)
were confirmed in the actual experiments at
COSY-J\"ulich using $Bi$ \cite{KUL98}, $Au$ \cite{KAM00} and $U$
targets \cite{KUL01}, where similar
cross sections could be observed experimentally.

Another important ingredient for the data analysis (to be
discussed later) is the velocity distribution of the hypernuclei
in the laboratory. The latter is dominantly determined by the
nucleon-nucleon and hyperon-nucleon cross section in the initial
stage of the reaction as modeled by the CBUU approach. It has been
shown in comparison to independent experimental data from refs.
\cite{FRA90,KOT74} that the momentum transfer to the residual nucleus
is well described by the transport approach in $p+U$ reactions
for $T_{lab}=$ 0.5 -- 3 GeV \cite{KUL01}.

\begin{figure}[h]
\resizebox{0.45\textwidth}{!}{%
  \includegraphics{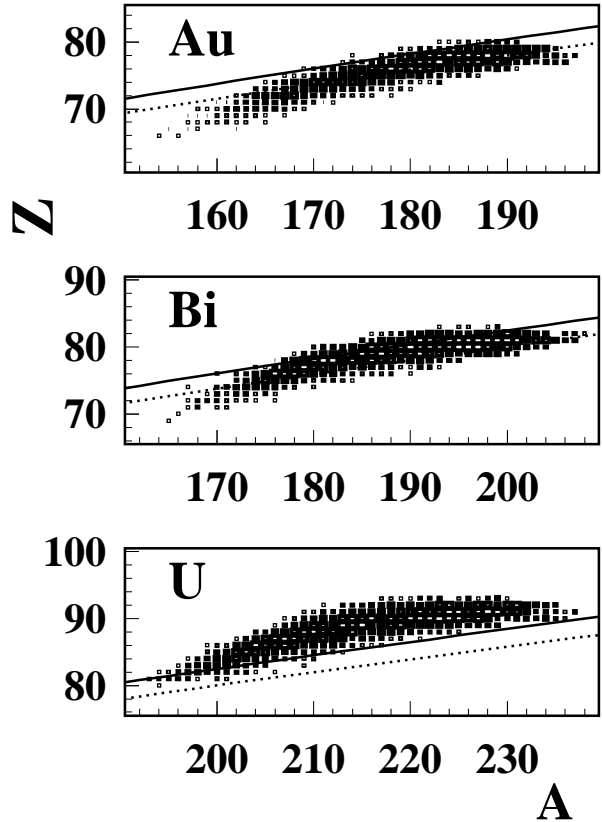}
}
\caption{ Two dimensional spectra  from CBUU +
evaporation calculations in charge $Z$ and mass $A$ of hypernuclei
for p + $^{197}$Au at T$_p$=1.7 GeV,
 p + $^{209}$Bi at T$_p$=1.9 GeV
and  p + $^{238}$U at T$_p$=1.9 GeV. The solid and dashed lines
 indicate hypernuclei  of fissility $Z^2/A =$ 34 and 32,
respectively. Delayed fission events essentially stem from nuclei
with fissility parameter ${Z^2}/{A}\geq 34$. } \label{fig:za}
\end{figure}

During the statistical decay phase the hypernucleus velocity
distributions cha\-nge only moderately, however,  a very
pronounced change in the mass and charge distributions is observed
\cite{RUD96}. The final charges and masses of
'cold' hypernuclei are correlated to form a valley of stability.
The resulting two-dimensional spectra in charge $Z$ and
mass $A$ of cold hypernuclei (typically after 10$^{-18}$ s) are
shown in fig. \ref{fig:za} in terms of cluster plots. These
differential distributions represent CBUU + evaporation model
calculations for hypernuclei produced in the reactions p +
$^{197}$Au at T$_p$=1.7 GeV, p + $^{209}$Bi at T$_p$=1.9 GeV,  p +
$^{238}$U at T$_p$=1.9 GeV. It is seen that the two dimensional
plots are quite similar for the three reactions considered, but
shifted in mass and charge according to the initial target. It
should be noted, that the width of the distribution in charge $Z$
remains rather constant as a function of mass $A$. This can be
inferred directly from the isospin-independent emission of protons
and neutrons in the pre\-equilibrium CBUU collision stage before the
Coulomb barrier is formed; after that the proton emission is
suppressed by the Coulomb barrier and neutron emission fills out
the 'valley of stability'.


It has to be pointed out that although the distributions in mass
differ by about 10 to 30 units for the different targets, they
have some common overlap region in the tails. Furthermore, the
$\Lambda$ induced fission probability essentially depends on the
fissility parameter  $Z^2/A$ (see Fig.3). The solid and dotted
lines in Fig. 3 show hypernuclei of $Z^2/A =$ 34 and 32,
respectively. We recall that only a fraction of the $(A,Z)$
distributions of hypernuclei created in $p+A$ interactions lead to
actual delayed fission events (see P$_f{_\Lambda}$ in Fig. 2),
i.e. essentially for $Z^2/A \geq$ 34. When averaging over the
experimental results for all targets one thus obtains a value for
$\tau_\Lambda$ that corresponds to an average over all nuclei with
masses A $\geq$ 180.


\section{Experimental setup and data analysis}

Hypernuclei produced in proton-nucleus collisions, which
survive the prompt fission stage, leave the target with a recoil
velocity $v_R$.
They subsequently decay at some distance from the
target proportional to the lifetime $\tau_\Lambda$ of the
$\Lambda$--hyperon and to the velocity $v_R$. Thus prompt and
delayed fission events can be separated by the spatial
distribution of their decays. The problem, however, is that the
prompt fission events are more frequent than the delayed fission
processes by factors of $\sim 10^5$ (cf. Fig. 2) --
which corresponds to the ratio of prompt
to delayed fission cross sections -- and the spatial
distribution of delayed events has to be measured with high
accuracy. The particular solution to this problem is provided by
the recoil shadow method \cite{MET74},  which allows to analyse
the spatial distribution of delayed decays with respect to the
product $\tau_\Lambda \cdot v_R$ in the presence of a huge background
compared to the investigated effect.


\begin{figure}
\resizebox{0.5\textwidth}{!}{%
  \includegraphics{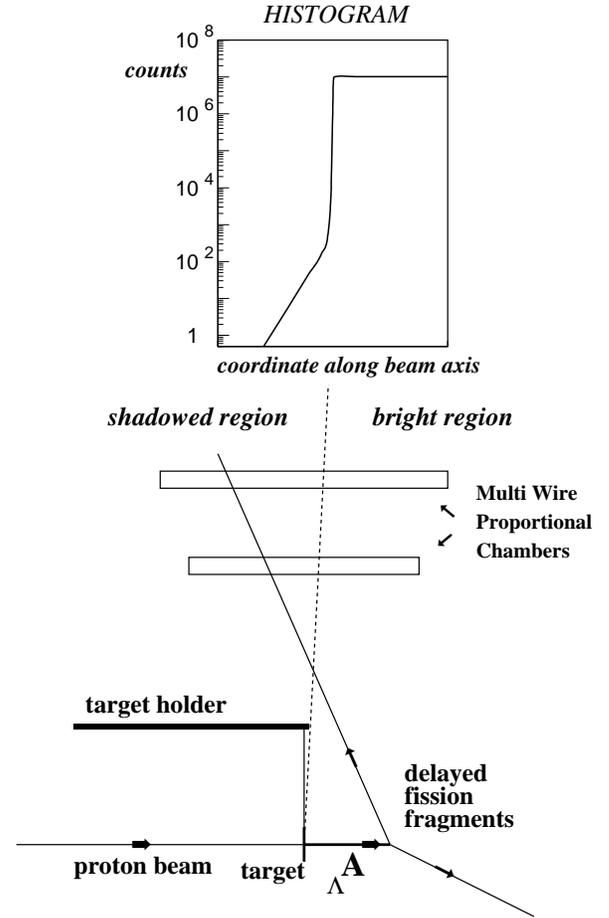}}
%
\caption{Schematic view of the experimental setup
and illustration of the recoil distance method (see text). The
dimension of the target holder and target in the lower part are
increased by a factor $\approx$ 30 relative to the MWPC's.}
\label{fig:fig1m}       
\end{figure}


A schematic view of the detection scheme \cite{PYS99} is shown in Fig.
\ref{fig:fig1m}, where the dimensions of the target and its holder
-- serving as a diaphragm -- are increased by a factor $\approx$
30  in comparison to the dimensions of the low
pressure multiwire proportional chambers (MWPC)
placed 30 cm from the target in a
direction perpendicular to the target.  The multiwire chambers are
sensitive to fission fragments, but not sensitive to protons and
other lighter particles. These detectors were partly screened by
the target holder such that the prompt fission fragments -
originating from the target - could not hit the shadowed (left)
part of the detector. This was, however, possible for fragments
from the delayed fission of hypernuclei $_\Lambda A$ escaping from
the target downstream the beam and fissioning in some distance
from the target. A schematic event distribution - projected on to
the beam axis - is shown in the upper part of Fig.
\ref{fig:fig1m}, which is characterized by an exponential fall-off
for the delayed fission events in the shadowed (left) region and a
constant (prompt) yield in the bright (right) region of the
detector.  For further details we refer the reader to ref. \cite{PYS99}.

In order to check whether the events detected in the shadow region
are not light particles or even $\gamma'$s, the following tests
have been performed:
\begin{itemize}
\item
The MWPC were irradiated with minimum ionizing particles
($\gamma$'s and $e^-$); it was shown that the detection
efficiency for such particles is below 10$^{-11}$.
\item
A pure carbon foil was used as a target in $ p+A$ measurements,
leaving the detection system unchanged. The measured spectra in
the shadowed part of the detectors were found to contain no
events.
\item
A $^{252}$Cf source was placed at the target position and a two--dimensional
energy loss versus time-of flight spectrum (between both MWPC)
was measured. The spectrum was populated in line with Monte Carlo calculations
taking into account the mass, charge and velocity distributions of fragments
according to the Viola systematics \cite{VIO85}
from the fission of californium.
\end{itemize}

The fragments, that hit the shadowed (left) part of the detector,
thus originate either from the delayed fission of hypernuclei (or
hyperfragments) or they are emitted in prompt fission from the
target and due to scattering on the shadow edge (part of the
target holder)  have changed their initial trajectories.
Therefore, scattering  creates a background in the shadow region
with an intensity proportional to the prompt fission cross
section. In order to determine the background distribution of hits
in the shadowed part of the detector measurements have been
performed at a much lower proton energy (T$_p$=1.0 GeV), where the
cross section for hypernucleus production is expected to be
negligibly small (about 4 orders of magnitude smaller than at 1.9
GeV), whereas the prompt fission yield is about the same.

 It has been shown by Monte Carlo simulations that  hyperfragments
from prompt fission of hypernuclei, that have changed their
direction due to the recoil induced by a subsequent
$\Lambda$-hyperon decay, can hit the shadowed region of the
detectors only in a very narrow region of 1-2 mm close to the edge
of the shadow region and, thus, do not contribute to the
distribution that was actually used for the extraction of the
lifetime of hypernuclei (see below).

The proton beam (with typically 5 $\cdot 10^{10}$ protons in the
COSY-ring) has been accelerated up to 1.9 GeV (for the observation
of hypernucleus production) and to 1.0 GeV (for an estimation of
the background originating from scattered fragments from prompt
fission of the target nucleus). The COSY accelerator was operated
in the supercycle mode, {\it i.e.} there were three cycles (each
of $\sim$15 s duration) of beam acceleration and irradiation of
the target; two of them  at the higher energy of 1.9 GeV and one
at 1.0 GeV.  This allowed to study the effect and the background
concurrently for the same shape and thickness of the target.


\begin{figure}
\resizebox{0.45\textwidth}{!}{%
\includegraphics{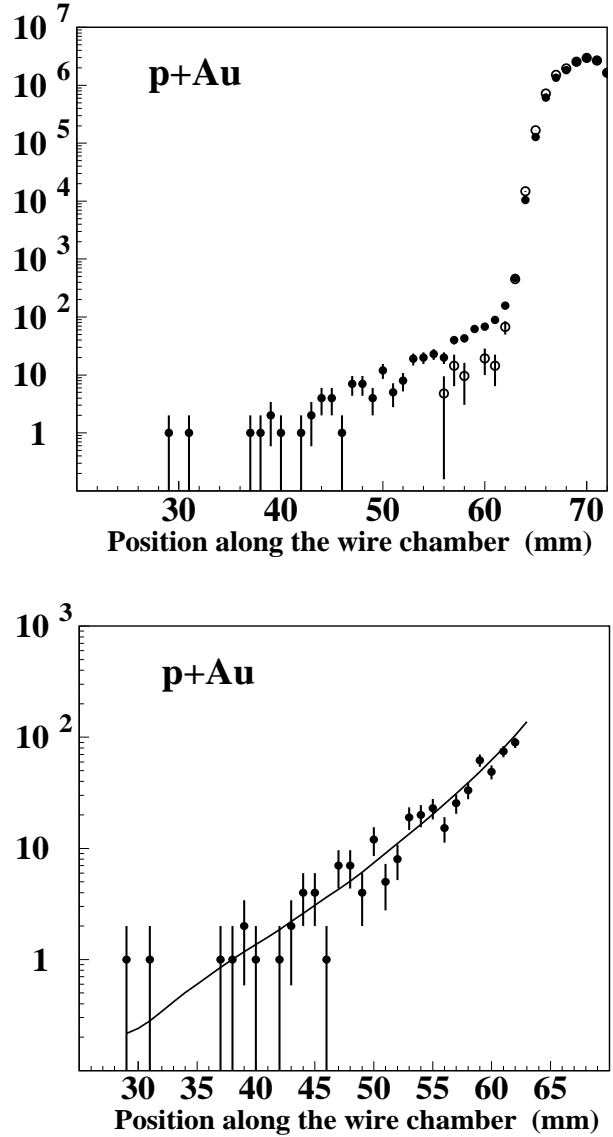}
}
\caption{Upper part: The position distribution of
hits of fission fragments in the position sensitive detectors for
the $p+Au$ experiment (from ref. \protect\cite{KAM00}). The full
dots represent the data for T$_p$=1.9 GeV whereas the open circles
show the data for T$_p$=1.0 GeV renormalized in the bright part of
the detectors to the 1.9 GeV data. Lower part: The position
distribution of hits from the delayed fission fragments of
hypernuclei in the shadow region obtained by subtracting the
background (renormalized data taken at 1.0 GeV) from the data
measured at 1.9 GeV. The solid line shows the result of the
simulation with the extracted value for the lifetime according to
the maximum likelihood method.} \label{fig:au}
\end{figure}
%


\begin{figure}
\resizebox{0.45\textwidth}{!}{%
\includegraphics{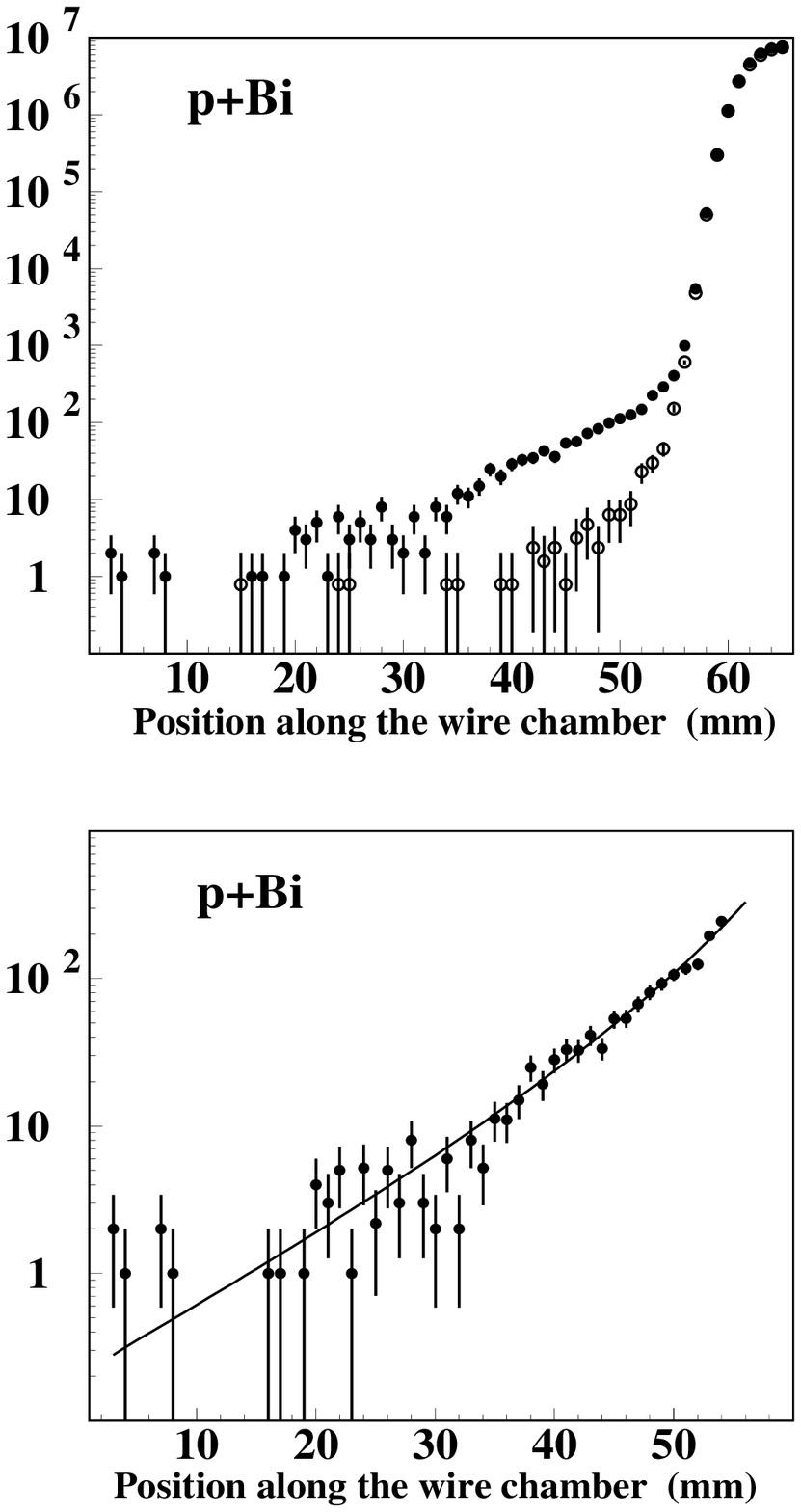}
}
\caption{The same distributions as in Fig.
\ref{fig:au}  for the $p+Bi$ experiment (from ref.
\protect\cite{KUL98}). } \label{fig:bi}
\end{figure}


\begin{figure}
\resizebox{0.45\textwidth}{!}{%
\includegraphics{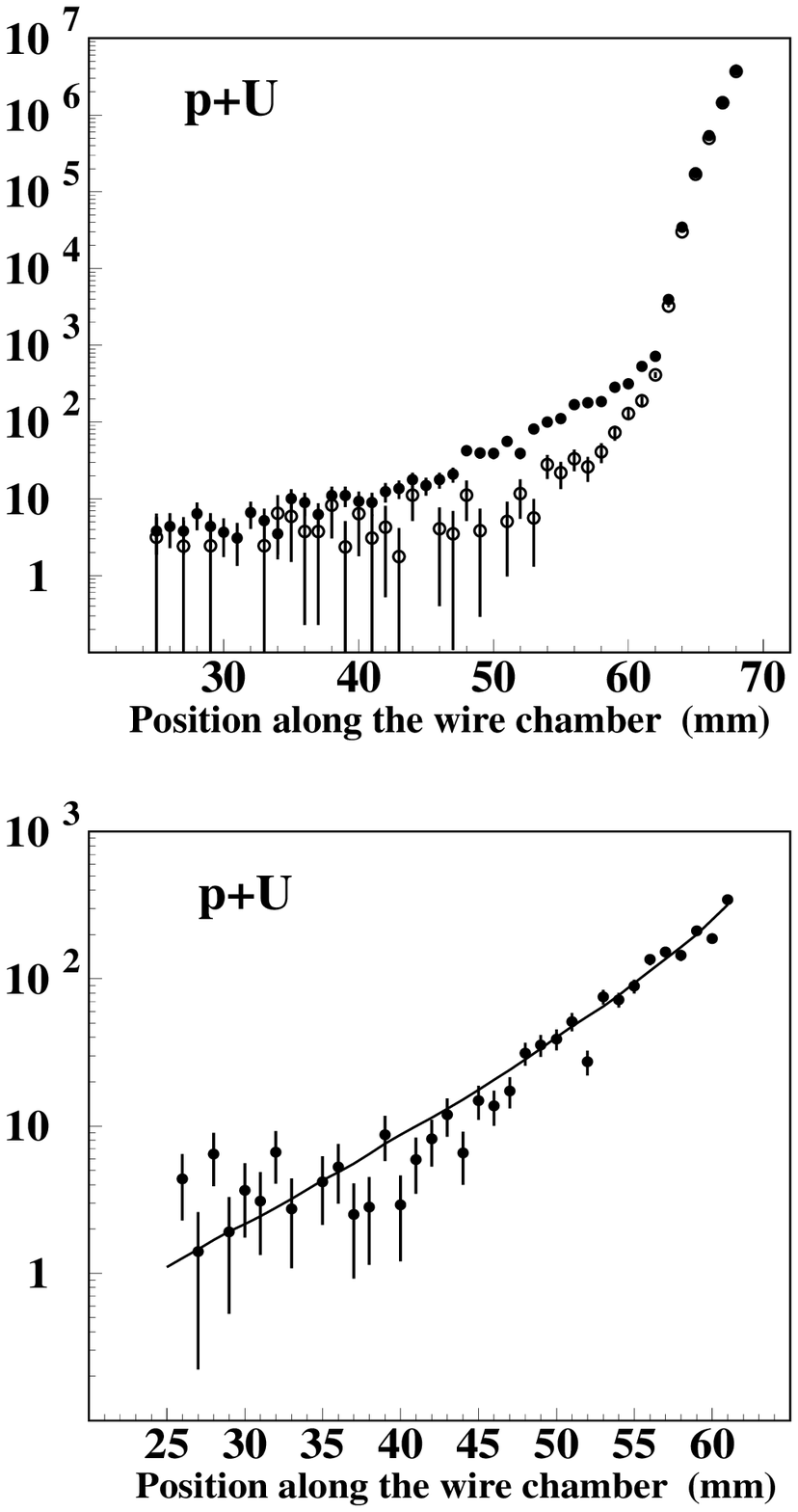}
}
\caption{The same distributions as in Fig.
\ref{fig:au}  for the $p+U$ experiment (taken from ref.
\protect\cite{KUL01}). } \label{fig:u}
\end{figure}
%


The distribution of hit positions of the fission fragments on the
surface of the detector then were projected on to the beam
direction. The respective distributions for the $Au, Bi,$ and $U$
target are shown in the upper parts of Figs. \ref{fig:au}, \ref{fig:bi},
\ref{fig:u} at
$T_p$ = 1.9 GeV (full dots) together with the background measured at
$T_p$ = 1.0 GeV (open circles).

These experimental distributions then have been compared with
simulated distributions, which were evaluated assuming the
velocity distribution of the hypernuclei (as obtained from the
CBUU + Hauser-Feshbach calculations) and a lifetime of the
$\Lambda$ - hyperon in the hypernuclei, where the latter was
treated as a free parameter in the fit procedure. Since the number
of events in the position distributions was not very large in some
experiments, a Poisson instead of Gaussian probability
distribution $p(n_i)$ has been used to simulate the number of
counts $n_i$ for each position bin (cf. ref. \cite{KUL01}). Then
the best lifetime $\tau_\Lambda$ was searched for by the  'maximum
likelihood'  method, which allows also for an estimate of the
statistical error for $\tau_\Lambda$ (see {\it e.g.} ref.
\cite{PAR00}). The results of the fits are shown in the lower part
of Figs. 5, 6, 7 by the solid lines in the shadow region in
comparison to the experimental data, where the background
(measured at $T_p$ = 1.0 GeV) has been subtracted from the 1.9 GeV
data.

The question arises, whether the velocity distributions of
hypernuclei might differ significantly when varying ($A,Z$).  In
such a case the simulation of the position distributions, which is
the crucial part in the analysis of the experimental data, should
be carried out by folding the velocity distributions of
hypernuclei with specified ($A,Z$) with the fission time
distributions of these hypernuclei. However, as detailed
calculations have shown \cite{RUD98}, those hypernuclei, that lead
finally to fission, have practically the same velocity
distribution.

\section{Summary of experimental results and error analysis}

\label{sec:discu}

In this section we summarize the  results of the COSY--13
Collaboration and compare to the lifetimes measured before (cf.
Table 1).  Such a comparison must necessarily involve a discussion
of experimental uncertainties. Whereas the {\it statistical}\, errors
can be unambiguously determined by the maximum likelihood
method  as described in detail in ref. \cite{KUL01}, the estimation
of {\it systematic}  errors has to be discussed individually for
each experiment, since the number of events in the shadow region of the
detectors have been different as well as the stability of the
individual targets during the irradiation periods.

\subsection{Systematic errors}

\label{sec:syste}

The systematic errors arise from:
\begin{description}
\item[ a.) ] the velocity distribution of hypernuclei,
\item[ b.) ] an anisotropic emission of the fission fragments,
\item[ c.) ] a nonuniform irradiation of the target by the proton beam,
\item[ d.) ] a change of position and shape of targets
 during the measurements,
\item[ e.) ] the background treatment in case of low statistics, and
\item[ f.) ] the explicit search procedure ($\chi^2$ or maximum likehood methods)
for the best lifetime.
\end{description}

Detailed simulations have been carried out to determine the
variation in the lifetime $\tau_\Lambda$ according to the error
sources listed above. The results of these studies in ref.
\cite{KUL01} lead to the actual numbers shown in Table 2 for the three
targets separately.


\begin{table}
\caption{The sources of systematic errors in the COSY-13
experiments. The total systematic error has been evaluated
assuming that the sign of all contributions is the same.}
\label{table:sources}
\begin{tabular}{llll}
 Source of errors &
Au & Bi &
U \\
\noalign{\smallskip}\hline a.) velocity distribution & 2 ps & 2 ps
&
2 ps \\
\noalign{\smallskip}\hline b.) anisotropic emission & 2 ps & 2 ps
&
2 ps \\
\noalign{\smallskip}\hline c.) nonuniform irradiation  & 4 ps & 4
ps &
4 ps \\
\noalign{\smallskip}\hline d.) change of shape and position & 2 ps
& 1 ps &
4 ps \\
\noalign{\smallskip}\hline e.) background treatment & 3 ps & 3 ps
&
3 ps \\
\noalign{\smallskip}\hline f.) search procedure & 2 ps & 2 ps &
2 ps \\
\noalign{\smallskip}\hline
Total & 15 ps & 14 ps &
17 ps \\
\end{tabular}
\end{table}


The systematic errors can be summed up to 15 ps for the Au target,
to 14 ps for the Bi target, and to 17 ps for the U target.

\subsection{Results}

We recall that due to the rather large dispersion in the ($A,Z$)
distribution of cold hypernuclei (cf. Fig. 3) the observation of
the delayed fission of these nuclei does not give an information
on the lifetime of specific heavy hypernuclei, i.e. with fixed
atomic number $Z$  and mass $A$, but it rather provides a lifetime
averaged over a group of different hypernuclei.


\begin{table}[h]
\caption{ The lifetime of heavy hypernuclei measured at
COSY-J\"ulich by COSY--13. The errors in the third column have
been obtained by quadratically adding the statistical and
systematic errors from the second column.  } \label{tab:2}
\begin{tabular}{lcll}
\hline\noalign{\smallskip} Target &  $\tau_{\Lambda}$ / ps &
$\tau_{\Lambda}$ / ps & Ref.
\\ \hline\noalign{\smallskip} Au  &
130$\pm$13(stat.)$\pm$15(syst.)& 130 $\pm$ 20 & \cite{KAM00} \\
\hline\noalign{\smallskip} Bi &  161 $\pm$7(stat.)$\pm$14(syst.)&
161 $\pm$ 16 & \cite{KUL98} \\
\hline\noalign{\smallskip} U  & 138
$\pm$6(stat.)$\pm$17(syst.)& 138 $\pm$ 18 & \cite{KUL01} \\
\hline\noalign{\smallskip}
\end{tabular}
\end{table}


A summary for the lifetimes $\tau_\Lambda$ including the
statistical and systematic errors is presented in Table 3. It can
be concluded, that the experiments performed with the proton beam
on $Au, Bi,$ and $U$ targets give consistent and comparable values
for the lifetime of heavy hypernuclei. Within statistics these
values are identical, though the average masses of the fissioning
hypernuclei differ for the different targets (cf. Fig. 3). On the
other hand, the individual distributions in ($A,Z$) overlap such
that we may also average over the three experiments to obtain an
average lifetime for hypernuclei with masses A $\approx$ 180 --
225 with a dispersion in charge $\Delta Z \approx$ 3 (for fixed A)
as:

\hspace*{2.5cm} $\tau_{\Lambda} = 145 \pm$ 11 ps  \hspace{0.5cm} (for $p$+A).


\begin{figure}[h]
\resizebox{0.5\textwidth}{!}{%
\includegraphics{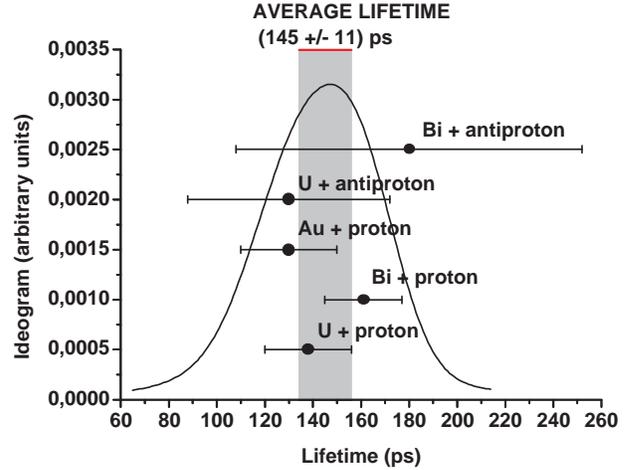}}
\caption{The lifetimes for proton and antiproton produced
hypernuclei on Au, Bi and U targets.  The horizontal bars present
the statistical and systematic errors added in quadrature. The
gray vertical bar displays the overall average value for the
lifetime of heavy hypernuclei and its width shows the error.  The
smooth Gaussian-like curve was evaluated as proposed in the Review
of Particle Physics \protect\cite{PAR00}, {\it i.e.} adding
Gaussian curves representing results from individual experiments.
Parameters of these Gaussian curves (average value and standard
deviation) are equal to the individual lifetimes and their errors
(square roots from sum of squares of statistical and systematic
errors).  The weights -- with which the individual curves enter
the sum -- were chosen as reciprocals of the errors quoted above.}
\label{fig:compi}
\end{figure}


\subsection{Comparison with antiproton induced reactions}

This average lifetime of heavy hypernuclei is within the
statistical error limits in agreement with the lifetimes
extracted from antiproton experiments Refs. \cite{ARM93} (see
Table \ref{tab:1}), which by averaging over the $Bi$ and $U$
targets amounts to:

\hspace*{2.5cm} $\tau_{\Lambda} = 143 \pm$ 36 ps \hspace{0.5cm} (for $\overline{p}$
+A).

\noindent In fact, the mass and charge distribution of hypernuclei
from the experiments with antiprotons should be similar to those
of the proton induced reactions since a comparable energy is
transferred to the nucleus. However, the latter reactions lead to
a much more precise value for $\tau_\Lambda$ since i) the
background can be determined experimentally in contrast to the
$\bar{p}$ induced reactions - which reduces the systematic errors
-- and ii) the velocity of the hypernuclei is much larger in the
laboratory due to the higher momentum transfer from the proton at
$T_p$ = 1.9 GeV. The latter fact also leads to a cleaner
separation of delayed fission events from prompt fission events in
the shadowed region of the detector, that stem from small angle
scattering in the target holder. Moreover, the geometrical
conditions of proton induced reactions allow for a less ambiguous
interpretation of fission fragment distributions in the shadowed
regions of the detectors than those for antiproton experiments
because in the former investigations it was possible to neglect
the contribution of hyperfragments which originate from prompt
(not delayed) fission of hypernuclei, but are observed in the
shadow region due to recoil caused by subsequent decay of the
$\Lambda$-hyperon.

A compilation of all results discussed above for the lifetime
$\tau_\Lambda$ from proton and antiproton induced reactions is
presented in Fig. \ref{fig:compi} in the form proposed in the
Review of Particle Physics \cite{PAR00}. We note, that adding the
result from the $\bar{p}$ experiments to the data from COSY--13 does
not change the number of $\tau_\Lambda = 145 \pm 11$ ps quoted
above.
%
%

\section{Implications for the $\Lambda N \rightarrow NN$ reaction}
\label{sec:impli}

\begin{figure}[h]
  \resizebox{0.38\textwidth}{!}{%
    \includegraphics{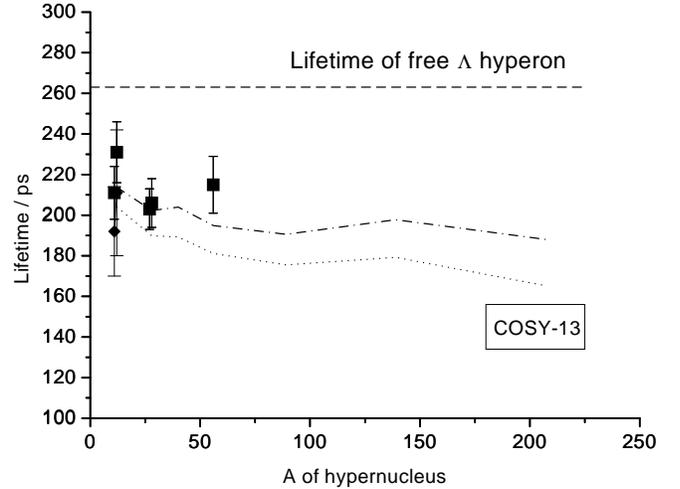}} 
    \caption{Mass dependence of the $\Lambda$-lifetime. Diamonds
    and squares represent experimental data obtained for light
    hypernuclei in refs \cite{GRA85,SZY91} and
    \cite{BHA98,PAR00A}, respectively.
    The rectangle placed at A $\sim$ 200 represents the experimental result obtained
    for heavy hypernuclei by the COSY-13 collaboration.
    The width of the rectangle indicates
    the range of masses of the hypernuclei observed in the COSY-13 experiments
    (180 $\le$ A $\le$ 225), whereas the height corresponds to the
    experimental accuracy quoted (11 ps).
    The dot-dashed line and dotted line present results of
    theoretical calculations from refs. \cite{ALB00A}
    and \cite{JID01}, respectively, while the
    horizontal
    dashed line shows the experimental lifetime of the free $\Lambda$ hyperon.
    }
 \label{fig:exptheo}
 \end{figure}

The dependence of the lifetime on the mass number of the
hypernucleus is shown in fig. \ref{fig:exptheo}.  The experimental
results for light hypernuclei (mass number 11 $\le$ A $\le$ 56)
seem to be mass independent within the limits of errors. The
lifetimes of heavy hypernuclei as measured by the COSY-13
collaboration  do not indicate a mass dependence in the studied
range of  mass numbers (180 $\le$ A $\le$ 225) either. However,
experiments show that the lifetimes of heavy hypernuclei are
shorter by $\sim$ 60 -- 70 ps than those for light hypernuclei.
This difference implies that the lifetime should decrease by less
than 0.5 ps per mass unit. Such a weak decrease could not be
established within the present experimental accuracy if the light
or the heavy hypernuclei are studied separately. In both cases the
covered mass range is  about 45 mass units  corresponding to a
variation of the lifetime by about 20 ps, i.e. less than two
experimental errors.


The dotted and dot-dashed lines in fig. \ref{fig:exptheo}
represent theoretical model expectations evaluated within the
meson exchange model.  In both calculations the validity of the
$\Delta$I =1/2 rule has been assumed and the contribution from
nonmesonic decays initiated by two nucleons was included.
Nevertheless, the results of the calculations differ rather
significantly, i.e. $\sim$10 ps for light and $\sim$20 ps for
heavy hypernuclei.

However, the smooth decrease of the lifetime versus mass of the
hypernuclei is a common property of both model calculations, i.e.
both approaches reproduce  qualitatively the mass dependence of
the experimental data as extracted from the comparison of
lifetimes of light and heavy hypernuclei. Furthermore, both models
predict a weaker decrease of the lifetime with mass than observed
in the experiment. The calculations of Alberico et al.
\cite{ALB00A} lead to a difference between the largest lifetime
(for ${}_\Lambda ^{12} C$) and the smallest one (for ${}_\Lambda
^{208} Pb$) of about 26 ps. Similarly, this difference in the
model of Jido et al. \cite{JID01} is  39 ps, whereas the
difference in the average experimental lifetimes from light and
heavy hypernuclei is approximately  60 -- 70 ps.  Such a large
discrepancy between theory and experiment should indicate an
inadequacy of one or more  model assumptions.

As discussed in the introduction and demonstrated in ref.
\cite{RUD99}, the $\Delta$ I = 1/2 rule might be violated in the
$\Lambda N \rightarrow NN$ interaction contrary to the case of
free hyperon decays. In this respect we recall that the lifetime
of heavy hypernuclei is sensitive to the ratio R$_n$/R$_p$ of the
neutron induced to proton induced $\Lambda$ nonmesonic decays
$\Lambda + N \rightarrow N + N$, whereas the lifetime of light
hypernuclei ($A \approx$ 12) is independent of this ratio. Thus, a
precise knowledge of the lifetime of light hypernuclei (which
depends only on R$_n$+R$_p$) and an accurate knowledge of the
lifetime of heavy hypernuclei (depending both on R$_n$+R$_p$ and
on R$_n$/R$_p$) enables us to determine the absolute
normalization, i.e. R$_n$+R$_p$, as well as the ratio R$_n$/R$_p$.

Furthermore, we can test the validity of the phenomenological
$\Delta$I = 1/2 rule due to the following reasons: The ratio
R$_n$/R$_p$ vanishes for final state isospin I$_f$=0 since the
neutron induced $\Lambda$ decay leads only to neutron-neutron
final states, which cannot form an isospin zero state. On the
other hand, the ratio R$_n$/R$_p$ is equal 2 for $\Delta$I = 1/2
decays to pure I$_f$=1 final states (realized e.g. for $\Lambda$ -
nucleon spin state $S$) \cite{SCH92}. Therefore, in the general
situation - where the observed decays correspond to an incoherent
mixture of the I$_f$=0 and I$_f$=1 final states - pure $\Delta$I =
1/2 decays must always result in a ratio R$_n$/R$_p \le$ 2. Any
measured ratio R$_n$/R$_p \ge$ 2 then will indicate a violation of
this rule. We will argue in the following that this should be
indeed the case.

\begin{figure}[h]
 \resizebox{0.45\textwidth}{!}
       {%
  \includegraphics{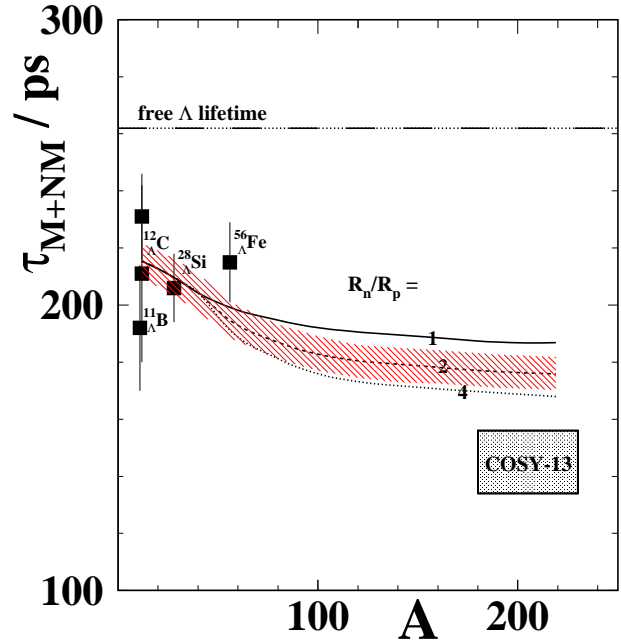}
       }
\caption{Calculations of the $\Lambda$-lifetime $\tau_{M+NM}$ due
to the mesonic and nonmesonic decay  as a function of the
hypernucleus mass $A$ in the valley of stability (from ref.
\protect\cite{RUD99}) in comparison to the data of Refs.
\protect\cite{BHA98,SZY91}.  The COSY--13 collaboration result for
nuclei with masses $A \ge$  180 is marked by the hatched area
labelled "COSY--13". The width and height of this rectangle
represent the range of hypernucleus masses involved and the error
of the lifetime determination, respectively.  In the theoretical
calculations both mesonic and non-mesonic decay modes are taken
into account whereas the  unknown ratio of the weak decay rates
$R_{n}$/$R_{p}$ is treated as a parameter with values:
$R_{n}$/$R_{p}$ =  1,2,4.  The hatched area around the dashed line
(corresponding to $R_{n}$/$R_{p}$=2) shows the $\pm \sigma$
uncertainty  in the magnitude of the weak transition
$\sim(R_{n}+R_{p})$ determined from the lifetimes  of light
hypernuclei with A$\approx$ 12.  } \label{final}
\end{figure}

To sharpen the arguments we show again the theoretical
calculations from ref. \cite{RUD99} for the $\Lambda$ hyperon
lifetime for both, the mesonic and nonmesonic contributions
included, in Fig. \ref{final} as a function of the hypernucleus
mass $A$. In these calculations the strength of the weak
transition $\Lambda N \rightarrow NN$ $\sim R_n + R_p$ is fixed in
magnitude  to the data (cf Fig. \ref{final}) for light hypernuclei
with $N \approx Z$ and masses $A$  $\approx$12. We mention that this
strength has an error of about 5\% according to a statistical
analysis of the lifetimes for these nuclei which amounts to $\approx \pm$7 ps
for heavy hypernuclei (A$\sim$200).

The calculations for a constant ratio R$_n$/R$_p$ then lead to a
smooth decrease for the lifetime as a function of mass $A$ which
approximately saturates for $A \approx$ 160 (solid line for
R$_n$/R$_p$=1). When increasing the ratio to R$_n$/R$_p =2$ we
obtain the dashed line  which is the lowest limit for the
$\Delta$I = 1/2 rule to hold according to the argumentation
presented above. Any further increase of R$_n$/R$_p$ (dotted line)
leads to a steeper dependence of $\tau_\Lambda$ with mass $A$
since in neutron rich nuclei -- along the line of stability -- the
$n \Lambda \rightarrow nn$ channel becomes the dominant one.


When comparing the different theoretical lines with the lifetime
extracted from the present work for masses $A \ge$ 180 (hatched
area with COSY-13), we find that a ratio R$_n$/R$_p \le$ 2 is not
compatible with $\tau_\Lambda = 145 \pm$ 11 ps for the heavy
hypernuclei. Thus within the described scenario
the $\Delta$I=1/2 rule is violated.

The latter conclusion also holds, when the contribution of two
nucleon induced decays ($\Lambda$+n+p $\rightarrow$ n+n+p) is
taken into account since it was shown by Ramos et al.
\cite{RAM97}, that the yield of two nucleon induced decays of
$\Lambda$ hyperons is independent of the mass of the hypernucleus.
The presence of such a mass-independent contribution effects the
mass dependence of lifetimes in the same way as a decrease of the
R$_n$/R$_p$ ratio, i.e. it makes the mass dependence less steep.
Therefore, an experimental indication for a steeper mass
dependence relative to the theoretical result for the one nucleon
induced decay -- under the assumption of the validity of the
$\Delta$I = 1/2 rule -- becomes an even stronger argument for a
violation of this rule when two nucleon induced decays contribute.

Since this conclusion is based on experimental data for lifetimes
of light and heavy hypernuclei, which are biased by statistical
and systematic errors, the violation of the $\Delta$I = 1/2 rule
can only be stated with some confidence level $P_c < 1$. To
estimate this probability we followed the error analysis described
in Ref. \cite{RUD99} using the present average value for the
lifetime of heavy hypernuclei (cf. fig. \ref{final}) with the
error evaluated as a sum of statistical and systematic errors (11
ps). This leads to a confidence level $\approx$ 0.98;
an inclusion of the antiproton data from ref. \cite{ARM93} (Table 1)
does not modify this result.

It should be emphasized, that the mass dependence
of the lifetime varies only weakly with the ratio  R$_n$/R$_p$
for large values of this ratio.
Thus the error in the  normalization of the
theoretical curves in fig. \ref{final}, \emph{i.e.} $\pm$ 7 ps --
the error of R$_n$+R$_p$  determined by the accuracy
of the lifetimes of light hypernuclei --
and the error of the lifetime for heavy hypernuclei,
 \emph{i.e.} $\pm$11 ps  (as extracted from the COSY--13 data)
do not allow to establish the ratio R$_n$/R$_p$
more precisely; it can only be stated that it is larger than 2.

\section{Discussion}

The conclusions presented above rely on: i) the accuracy of the
overall normalization, which is a free parameter of the present
theoretical model, and ii) the assumption that the model
predictions with respect to the mass dependence of the hypernuclei
lifetimes are reliable. We will discuss these premises in the
following.

We recall that the theoretical model formulated in ref.
\cite{RUD99} is based on the transport Boltzmann-Uehling-Uhlenbeck
equation (BUU). It treats the nuclei as systems of fermions in a
selfconsistent mean field with mutual in-medium interactions
allowed by the Pauli principle. Due to the semiclassical limits
invoked the approach neglects the shell structure of the nuclei.

The decay width of the nonmesonic decay is evaluated from the
collision rate of hyperons with nucleons in the target using local
Thomas-Fermi distributions for the nucleon phase-space density and
a 1s state wavefunction for the hyperon. Since the cross section
for the $\Lambda$+N $\rightarrow$ N+N weak process is not known
from experiment, it was assumed in ref. \cite{RUD99} that the
differential cross section for the weak $\Lambda$+N $\rightarrow$
N+N process is proportional to the cross section of elastic
$\Lambda$+N $\rightarrow$ $\Lambda$+ N scattering. This is the
most far-reaching approximation of this model, which may influence
both, the normalization of the mass dependence of $\tau_{\Lambda}$
and the shape of this mass dependence.

ad i) The absolute normalization  -- a free parameter of the model
-- is responsible for all mass independent factors.  It has  been
determined from a comparison of the experimental and theoretical
results for light hypernuclei, where our model results do not
depend on the ratio R$_n$/R$_p$. In detail: The normalization has
been performed to an average value of the lifetimes for
$^{11}_{\Lambda}$B and $^{12}_{\Lambda}$C
\cite{PAR00,BHA98,SZY91}.  Within this normalization the
confidence level for a violation of the $\Delta$I=1/2 rule is
found to be $\sim$ 0.98. Taking the experimental lifetimes for
$^{12}_{\Lambda}$C and for $^{11}_{\Lambda}$B \cite{PAR00,SZY91}
separately gives confidence levels of 0.99 and 0.95 for a
normalization to  $^{12}_{\Lambda}$C and to $^{11}_{\Lambda}$B,
respectively. This shows, that the uncertainty in the
normalization cannot change the conclusion concerning the
violation of the $\Delta$I=1/2 rule.

ad ii)
However, as mentioned above, the lack of knowledge of the
elementary
cross section for the weak $\Lambda$+N $\rightarrow$ N+N process
might influence the shape of the mass dependence of
$\tau_{\Lambda}$.

To check the sensitivity of $\tau_{\Lambda}$ on the elementary
cross section for the weak $\Lambda$+N $\rightarrow$ N+N process
the calculations have been performed also with a constant (energy
independent) cross section and compared with the results of ref.
\cite{RUD99}. It was found that the mass dependence for the energy
independent cross section turned out to be even flatter in mass
$A$ and the lifetimes for heavy hypernuclei increased by $\sim$ 13
ps. Such a limit increases the difference between the experimental
lifetime of heavy hypernuclei and the theoretical model and thus
more strongly supports a violation of the $\Delta$I=1/2 rule.

Furthermore, to explore the least favorable situation for
rejecting the validity of the $\Delta$I=1/2 rule, i.e. assuming a
much steeper mass dependence, we have evaluated the confidence
level for the case, where the limiting curve for R$_n$/R$_p$=2 is
shifted downwards by 20 ps for heavy hypernuclei. Even for such a
significant  modification of the model the confidence level is
still quite large,   $\sim$0.75, in favor of a violation of the
$\Delta$I=1/2 rule.


 \begin{figure}[h]
  \resizebox{0.45\textwidth}{!}
        {%
  \includegraphics{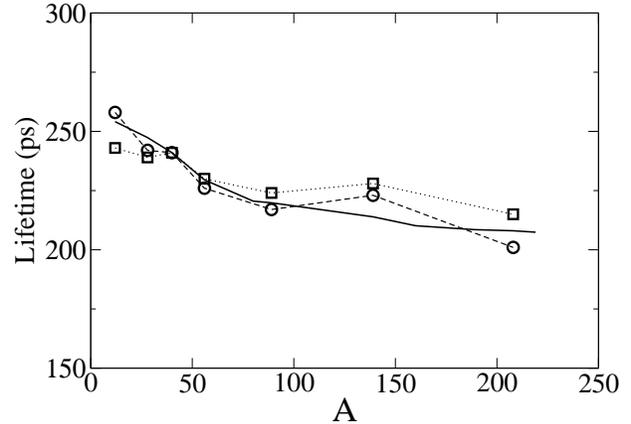}
        }
 \caption{Mass dependence of the $\Lambda$-lifetime
 $\tau_{M+NM}=1/(\Gamma_M+\Gamma_1)$ due to the mesonic and
 nonmesonic decay induced by single nucleons -- as evaluated in
 refs. \cite{ALB00A} (squares) and \cite{JID01} (circles) -- in comparison
 with the mass dependence calculated in ref. \cite{RUD99} for R$_n$/R$_p$=2
 and normalized to each other  at $A$=40.
 }
 \label{comp_th}
 \end{figure}


To gain further insight into the validity of our theoretical model
we have, furthermore, compared  the mass dependence of the
hypernucleus lifetime from ref. \cite{RUD99} with the mass
dependence from the more recent calculations of W. M. Alberico et
al. \cite{ALB00A} and D. Jido et al. \cite{JID01}. Since in both
studies the validity of the $\Delta$I=1/2 rule has been assumed,
we compare the mass dependence from ref. \cite{RUD99} within the
same assumption (i.e. the limiting value
 R$_n$/R$_p$=2 has been adopted) and omitted the
contribution of two-nucleon induced $\Lambda$-hyperon decays in
the results of refs. \cite{ALB00A} and \cite{JID01}, since the
model of ref. \cite{RUD99} does not include this contribution.

The calculated results for the mass dependence of $\tau_{\Lambda}$
from these three models are presented in Fig. \ref{comp_th}, where
the
squares correspond to the calculations of ref. \cite{ALB00A},
the circles to the calculations of ref. \cite{JID01} while the
solid line shows the mass dependence from ref. \cite{RUD99} after
normalization to the lifetime of $^{40}_{\Lambda}$Ca, which is
predicted by the two other studies to be exactly the same.  The
agreement of the
mass dependence from ref. \cite{RUD99} with the results
of the two other works is quite remarkable. In our opinion this
points towards a satisfactory reliability of the phase-space model
 \cite{RUD99}.

We thus conclude, that  in spite of the uncertainty in the shape
as well as  the uncertainty in the overall normalization of the
theoretical mass dependence of $\tau_{\Lambda}(A)$ the
experimental lifetime from COSY-13 is small enough to derive valid
conclusions concerning a violation of the $\Delta$I=1/2 rule.\\

Our conclusions -- which specify
that the ratio of neutron induced to proton induced
weak decays of the $\Lambda$ hyperons in heavy hypernuclei is larger than 2
-- should be confronted
with available  results obtained from experiments,
where the ratio $\Gamma_n/\Gamma_p$ was
measured by a straightforward detection of nucleons from the decay
of hypernuclei.
These results are shown in table \ref{tab:x}.

The data for heavy hypernuclei on this ratio have been obtained in
refs. \cite{LAG64,CUE67,GAN67} by using photographic emulsions to
observe the decays of heavy hypernuclei in the mass range  40 --
100. An analysis of the  energy spectra of fast protons was used
for this purpose. In all these works a dominance of neutron
induced over proton induced decays has been reported with a
$\Gamma_n/\Gamma_p$ ratio in the range from 1.5 to 9.0, which is
in line with our findings.

On the other hand, a much smaller ratio of $\Gamma_n/\Gamma_p$
($\sim$ 1) has been observed for light hypernuclei, where also
spectra of fast protons have been analysed. Here e.g. the results
by J.J. Szymanski at al. \cite{SZY91} for $^5_{\Lambda}He$,
$^{11}_{\Lambda}B$ and $^{12}_{\Lambda}C$ are smaller than 2; this
is also in line with the recent measurements of O. Hashimoto et
al. \cite{HAS02} for $^{12}_{\Lambda}C$ and $^{28}_{\Lambda}Si$.


\begin{table}[h]
\caption{The ratios of decay widths $\Gamma_n/\Gamma_p$ for light
and heavy hypernuclei obtained from a straightforward detection of
fast protons from the nonmesonic decay of the $\Lambda$ hyperon  }
\label{tab:x}       
\begin{tabular}{cll}
\hline
\noalign{\smallskip} Hypernucleus or range& $\Gamma_n/\Gamma_p$ & Ref.\\
 of masses of hypernuclei &                       &  \\
\noalign{\smallskip}\hline\noalign{\smallskip} $40 < A < 100$ &
1.5 -- 9.0 & \cite{LAG64}    \\
$40 < A < 100$ & 9.0 & \cite{CUE67}   \\
$ A \sim 50 $ & $\sim$ 5 & \cite{GAN67}  \\
$^{28}_{\Lambda}Si$ & 1.38$^{+0.13+0.27}_{-0.11-0.25}$ & \cite{HAS02}\\
$^{12}_{\Lambda}C$ & 1.33$^{+1.12}_{-0.81}$ & \cite{SZY91}\\
$^{12}_{\Lambda}C$ & 1.17$^{+0.09+0.020}_{-0.08-0.18}$ &\cite{HAS02}\\
$^{11}_{\Lambda}B$ & 1.04$^{+0.59}_{-0.48}$ & \cite{SZY91}\\
$^5_{\Lambda}He$ & 0.93$\pm$0.55 & \cite{SZY91}\\
\noalign{\smallskip}\hline
\end{tabular}
\end{table}


The experimental situation thus appears to create a puzzle;
$\Gamma_n/\Gamma_p$ is found to be larger than 2 for heavy
hypernuclei whereas it is apparently close to unity for light
hypernuclei. Thus, either the analysis of the experiments is
biased by some mass-dependent effects, or this ratio is indeed
different for light and heavy hypernuclei.

\section{Summary}

In this work we have summarized the experimental studies of the
COSY--13 Collaboration that aimed at measuring the lifetime
$\tau_\Lambda$ of the $\Lambda$ hyperon in heavy nuclei produced
in proton induced reactions
on $Au, Bi$ and $U$ targets employing the recoil shadow method.
The lifetimes extracted from the various experiments are all
compatible with each other and also with the  lifetimes
determined by early antiproton annihilation experiments
on $Bi$ and $U$ targets from
ref. \cite{ARM93}, however, much more accurate. These lifetimes
correspond to a broad range in mass and charge of the produced
hypernuclei  (cf.
Fig. 3) with a rather narrow dispersion in charge (for fixed A).
This mass range is comparable to the mass range of light hypernuclei
studied up to now.

Averaging the lifetime $\tau_\Lambda$ over all results from  the
COSY--13 measurements we obtain

\hspace*{2.5cm} $\tau_{\Lambda} = 145 \pm$ 11 ps.

This value for the lifetime of heavy hypernuclei is smaller than
the results of recent theoretical calculations by W.M. Alberico et
al. \cite{ALB00A} ($\sim$ 188 ps) and D. Jido et al. \cite{JID01}
($\sim$ 165 ps), which have been performed for the full range of
masses of hypernuclei, by more than 3 and 2 standard deviations
(in the first and the second case, respectively). In the framework
of the theoretical model of ref. \cite{RUD99} such a small value
for $\tau_{\Lambda}$ may be explained by a dominance of the
neutron induced over proton induced decay rates (R$_n$/R$_p>2 $).
This implies that the empirical $\Delta$I= 1/2 isospin rule --
found for the vacuum decays of single strange hadrons and assumed
to be valid in the theoretical calculations of W.M. Alberico et
al.\cite{ALB00A} and D. Jido et al.\cite{JID01}
 -- is violated
for the in-medium $\Lambda N \rightarrow NN$ transition.
The latter reactions involve a high momentum
transfer, \emph{i.e.} they test the $\Lambda N$ weak interaction at short
distances, where the overlap of the quark wave functions is very
large. It is questionable, if these compact 'parton
configurations' might be described properly in the meson-exchange
picture based on effective hadronic lagrangians. A description
with partonic degrees of freedom, which includes automatically
$\Delta$I=3/2 transitions, should be more adequate, but
reliable calculations on the partonic level still have to wait for
future.

\vspace*{1.5cm} {\bf Acknowledgements}

\noindent This work has been supported by the DLR International
Bureau of the BMBF, Bonn, and the Polish Committee for Scientific
Research.


\end{document}